\documentclass[aps,pre,onecolumn,epsf,graphicx,showpacs]{revtex4}

\def\bec{\begin{center}}
\def\enc{\end{center}}

\def\ben{\begin{equation}}
\def\ba{\begin{array}}
\def\bea{\begin{eqnarray}}

\def\een{\end{equation}}
\def\eea{\end{eqnarray}}
\def\ea{\end{array}}
\def\btab{\begin{table}}
\def\btabu{\begin{tabular}}
\def\etab{\end{table}}
\def\etabu{\end{tabular}}
\def\bit{\begin{itemize}}
\def\eit{\end{itemize}}
\def\bef{\begin{figure}[htb]}
\def\befh{\begin{figure}[!h!]}
\def\enf{\end{figure}}

\def\b1{{\bf 1}}

\def\cos{\hbox{cos}\:}
\def\sin{\hbox{sin}}

\usepackage{epsfig,latexsym}

\newcommand \bew {\begin{widetext}}
\newcommand \enw {\end{widetext}}

\begin{document}

\title{\bf\noindent Some observations on the renormalization of membrane 
rigidity by long-range interactions}

\author{D.S. Dean$^{(1)}$ and R.R. Horgan $^{(2)}$}

\affiliation{
(1) IRSAMC, Laboratoire de Physique Quantique, UMR CNRS 5152, Universit\'e Paul Sabatier, 118 route de Narbonne, 31062 Toulouse Cedex 04, France\\
(2)DAMTP, CMS, University of Cambridge, Cambridge, CB3 0WA, UK
}

\date{7th October  2005}

\begin{abstract}
We consider the renormalization of the bending and Gaussian rigidity of 
model membranes  induced by long-range interactions between the components
making up the membrane. In particular we analyze the effect
of a finite membrane thickness on the renormalization of the bending and 
Gaussian rigidity by long-range interactions. Particular attention 
is paid to the case where the interactions are of a van der Waals type.

\end{abstract}  
\maketitle
\vspace{.2cm}
\pagenumbering{arabic}

\pagestyle{plain}
\section{Introduction}
In the Monge gauge the simplest Helfrich Hamiltonian \cite{hel}
used to describe the 
bending energy of a membrane, whose height fluctuations in the direction 
perpendicular to the projection plane are written as $z = h({\bf x})$,
is given by 
\begin{equation}
H_{Hel} = {\kappa_b\over 2} \int_{A_0} d{\bf x} \left(\nabla^2 h({\bf x})\right)^2.\end{equation}
Here $\kappa_b$ is the bending rigidity and it is principally generated
by short range lipid-lipid interactions. Clearly however $\kappa_b$ will also
depend on long-range interactions between the membrane components such as 
van der Waals and electrostatic interactions. Unlike  short range interactions,
these long-range interactions can in some cases be modified rather easily, 
for example  one can screen electrostatic interactions by adding electrolyte 
to the system.

A number of authors have considered the problem of renormalization of bending
and Gaussian rigidities due to long-range interactions 
\cite{winh,lek,klein,kum,pinc,netz,deho}. Clearly this renormalization is 
important in determining membrane morphology and is thus of great physical 
and biological importance.  The first effects studied were those due
to the presence of a surface charge on the membrane surface, more
precisely an inner and outer surface charge. These systems were first studied
using the Poisson Boltzmann mean field theory in its linearized form
\cite{winh} and subsequently  in its full non-linear form \cite{lek,kum}. 
In the case of fixed symmetric surface charge (with respect to the two 
surfaces of the membrane)  it is found in these  studies 
that renormalization of the bending rigidity is positive, however the
renormalization  of the Gaussian rigidity is found to be negative. 
At the mean field level, even in the 
presence of a screening electrolyte solution, the effective interaction
between like charged surfaces is repulsive, the increase of  the bending
rigidity is thus easy to understand from a physical point of view. The 
reduction of  the Gaussian rigidity in these systems  is less obvious to 
understand from a more intuitive stand point. The general 
conclusion coming from the studies of charged membranes is 
that  the presence of surface charges can favor their buckling and  
perhaps induce the formation of spherical vesicles.

The literature on the renormalization of membrane rigidity by long-range
interactions contains a few contradictory results, some of which this paper 
will try to partially resolve. Firstly a number of authors predict 
that non retarded van der Waals interactions between the membrane 
components will lead to  a reduction of $\kappa_b$ \cite{netz,deho}. 
These non-retarded interactions 
are attractive  and the  general formalism developed by Netz 
\cite{netz} predicts  that attractive pair-wise inter-membrane 
interactions will  lead to a reduction of  $\kappa_b$. 
This seems eminently reasonable from a physical perspective,  
attractive interactions favor points on
the membrane becoming closer together and thus a flat sheet which is 
rolled into a cylinder should have a lower
energy. This reasoning is however rather deceptive, indeed we shall later 
see that while attractive forces favor the formation of a cylinder from a 
plane sheet, they may actually inhibit the formation of a spherical
vesicle.  

Taking a flat membrane and rolling it up  to form a cylindrical surface of 
length $L$ and radius $R$ costs a  bending free energy given by
\begin{equation}
H_b = {\kappa_b \pi L\over R} \label{eqfb}.
\end{equation}
As mentioned above, it seems clear that at a fixed surface area, 
if the interaction is attractive then it is energetically favorable 
to form a cylinder from a flat sheet if the only energies involved 
are due to attractive long-range interactions
between the membrane components. Hence, if this is the case, the 
renormalization of $\kappa_b$ above should be negative. 
In  a recent paper \cite{deho} we showed that this 
is indeed the case, in agreement with Netz \cite{netz} who used his approach 
for a generic pairwise potential applied to  the case of effective 
dipole-dipole interactions. 
In \cite{deho} the static van der Waals interactions are taken  into 
account via the  difference in the dielectric constant of the 
membrane of finite thickness and the surrounding bulk, {\em e.g.}
water, using the  Lifshitz theory \cite{lif}. If $\delta$ is
the membrane thickness, we showed that for large $R$ and when 
the dielectric constant of the membrane $\epsilon_M$ is close to the 
dielectric constant of water $\epsilon_W$,   the van der Waals
contribution to the free energy can indeed be seen as a renormalization
of the bending rigidity {\em i.e.} the free energy is given by Eq. (\ref{eqfb})
with  
\begin{equation}
\kappa_b = -{k_B T \over \pi}\frac{\Delta^2}{64}\left[3\log(\Lambda \delta) + 
0.02954\right], \label{vdw1}
\end{equation}
where $\Delta = (\epsilon_W - \epsilon_M)/(\epsilon_W + \epsilon_M)$
and $\Lambda \sim 1/a$ where $a$ is a microscopic cut-off scale.
In an older paper \cite{klein} the same system was analyzed and the 
resulting $\kappa_b$ was predicted to be positive, implying a membrane 
stiffening due to the van der Waals interactions. More precisely  in 
\cite{klein} it is predicted that, for large $R$,  
$H_b \sim Const.\times \ln(R\Lambda)/R$.
The difference between the results of \cite{deho} and \cite{klein}
can be traced to the fact that the expression for the free energy given in
\cite{klein} does not go to zero when the thickness $\delta$ of the 
membrane goes to zero, {\em i.e.} when no membrane is present. When the 
result of \cite{klein} is normalized to give zero when $\delta =0$ then 
the result of \cite{deho} is obtained. The apparent logarithmic behavior
in the result of  \cite{klein} is found to disappear, although the mechanism
is very subtle and involves some very lengthy analysis. The calculation of 
\cite{deho} also predicts that the renormalization of $\kappa_b$ is negative
for arbitrarily large differences in dielectric constants, {\em i.e.}
beyond the dilute approximation where the description in terms of an
effective  pairwise interaction is no longer valid.

An electro-neutral membrane of infinitesimal thickness containing 
monovalent charges, such as a 1-1 electrolyte,  has been studied in 
\cite{pinc}. In this study the surrounding medium is taken to be a non-ionic
solvent (a somewhat unlikely situation) . It was shown that for this system
a cylindrical geometry has a bending free energy due to the charges of
\begin{equation}
 H_b = -{k_B T L\over 24 R}\ln(R/\lambda_D), 
\end{equation}
here  $1/\lambda_D = 8\pi n_0 l_B$ where $l_B$ is the Bjerrum length and
$n_0$ is the (positive/negative) surface charge density of the membrane. 
This predicts an $R$ dependent bending rigidity
\begin{equation}
\kappa_b = -{k_B T \over 24 \pi  }\ln(R/\lambda_D).\label{spincus}
\end{equation}
Subsequently Netz analyzed this problem in his  
general formalism  where the area elements of the membrane interact 
via a generic pairwise interaction. In order to do this he was obliged to 
calculate and effective pairwise interaction for the system. 
He found that for this particular system 
\begin{equation}
\kappa_b = -{k_B T \over 384 \pi}.\label{snetz}
\end{equation}
It is interesting to ask what is the origin of these differences. In the 
method of \cite{pinc} one looks at the fluctuations about the mean field
and effectively calculates a functional determinant related to the surface 
charge fluctuations. In this calculation the energy of a flat membrane
is explicitly subtracted off to yield the bending energy. 
In the Netz formalism the  energy  is expanded to quadratic order in  small
height fluctuations about a flat membrane, the perturbation being taken about
the true area of the membrane as opposed to the projected area. 
Possible reasons for the
observed discrepancy, assuming both calculations to be formally correct, are
\begin{itemize}
\item There is an ensemble difference: the energy of a flat membrane is 
only explicitly subtracted off  in the calculation of \cite{pinc}
\item As pointed out by Netz \cite{netz},  his calculation is done as a 
perturbation about a flat membrane but the calculation of \cite{pinc} is 
explicitly done for a cylindrical (and  spherical geometry). This could lead to
a difference.
\item The effective interaction needs to be calculated in 
this system  and there could be a difference in the approximations used.

\end{itemize}

In what follows we shall show that the formalism of Netz can be applied 
with an explicit subtraction of a flat membrane energy of same area and that,
under rather weak assumptions, the same result is recovered. This cannot 
therefore be at the origin of the disagreement. In addition we show that 
for a general potential $V$ between membrane components a calculation 
done in a cylindrical geometry agrees with the result of Netz based on
a perturbation about a flat plane, however the potential
$V$ must be of sufficiently short range. This result makes sense as the 
low momentum height fluctuations correspond to regions of low and slowly 
varying curvature, and it is thus unlikely that this could lead to a 
difference, at least for short range potentials. 
We then use the geometric method of calculation to compute the 
Gaussian rigidity $\kappa_g$ by examining the case of a sphere.
In some cases the formulas for $\kappa_b$ and $\kappa_g$ 
obtained for an infinitesimally thin membrane need to be regularized by a
short distance cut-off in the potential $V$. However membranes have a finite
thickness and it is thus interesting to investigate the effect of a finite
membrane thickness to see if this regularizes the rigidity renormalizations.
The renormalization of rigidities for finite membrane thickness are 
therefore also derived via the geometric route. 

As an application we carry out 
the calculation of renormalization the bending and Gaussian rigidities
of a membrane of finite thickness $\delta$  having a  differing 
dielectric constant
to that of its surrounding medium..
This calculation is carried out in the dilute limit where the two
dielectric constants are close to each other and is equivalent to
computing the effect of the zero frequency van der Waals interactions.
In this calculation  the effective pairwise van der Waals interaction is 
regularized with a real space cut-off, as opposed to the Fourier space 
cut-off employed in the eigen-mode expansion of \cite{deho}. The result
of this calculation with a real space cut-off is strictly identical
to that of Eq. (\ref{vdw1}) up to the definition of the cut-off scale.
The contribution to the Gaussian rigidity is also computed and  
found to be positive and with  leading divergence in the  small cut-off, again
denoted by $a$, of the form $\delta^2/ a^2$.

\section{The Netz approach explicitly renormalized with respect  
to a flat membrane}

Here we revisit the approach of  Netz 
for the computation of the renormalization of the bending rigidity of a
close to planar membrane by long-range interactions. 
The argument we give is  slightly different 
in that we use a formalism which calculates the energy of a fluctuating
membrane with respect to that of a flat membrane with no fluctuations but
with the same surface area. Reassuringly this approach gives exactly the 
same result.  We imagine a membrane with projected area $A_0$
in the Monge gauge over the region ${\bf x}\in A_0$ and of height 
$h({\bf x})$ in the $z$ direction at that point. If the membrane is allowed to 
fluctuate in the $z$ direction 
then the bending energy due to these fluctuations is given by:
\begin{eqnarray}
H_b &=& {1\over 2}\int_{A_0\times A_0} d{\bf x} d{\bf x' }
\ \sqrt{1+ \left(\nabla h({\bf x})\right)^2}
\sqrt{1+ \left(\nabla h({\bf x'})\right)^2}
V\left(\sqrt{({\bf x}-{\bf x'})^2 + (h({\bf x}) - h({\bf x'}))^2}\right)
\nonumber \\
&-&{1\over 2}\int_{A\times A} d{\bf x} d{\bf x' }
V\left(\sqrt{({\bf x}-{\bf x'})^2}\right)\label{eqh1}
\end{eqnarray}

Note that the dimensions of $V$ as defined above 
are thus $[E]/[L^4]$, $E$ denoting energy and and $L$ length.
In the case where $V$ is a purely energetic interaction then $H_b$ is a 
purely energetic term 
However if $V$ is an effective interaction induced 
by thermodynamic effects, ({\em e.g.} presence of electrolyte when the membrane
is charged or static van der Waals interactions), then it will have a 
temperature dependence and $H_b$ is thus strictly speaking  a free energy.
The real area (as opposed to the projected area)  of the membrane is given by 
\begin{equation}
A = \int_{A_0}d{\bf x}\ \sqrt{1+ \left(\nabla h({\bf x})\right)^2}
\end{equation}
and the second term in Eq. (\ref{eqh1}) corresponds to the subtraction of  
the energy of a flat membrane of the same area, which
can be thought of as a flat bulk membrane from which the membrane we
study is drawn from. To quadratic order in the height fluctuations $h$ we
obtain 
\begin{eqnarray}
H &=& {1\over 2}\int_{A_0\times A_0} \ d{\bf x} d{\bf x' } V(|{\bf x} -{\bf x'}|)
- {1\over 2}\int_{A\times A} \ d{\bf x} d{\bf x' } V(|{\bf x} -{\bf x'}|)
\nonumber \\
&+&{1\over 2}\int_{A_0\times A_0} \ d{\bf x} d{\bf x' }\ V(|{\bf x} -{\bf x'}|)
\left(\nabla h({\bf x})\right)^2
+{1\over 4}\int_{A_0\times A_0} \ d{\bf x} d{\bf x' }\ {V'(|{\bf x} -{\bf x'}|)
\over |{\bf x} -{\bf x'}|}(h({\bf x}) - h({\bf x'}))^2,
\end{eqnarray}
the second term depends on $h$ through the area $A$ which to this 
quadratic order is given by
\begin{equation}
A = A_0 + {1\over 2} \int_{A_0} d{\bf x} \left(\nabla h({\bf x})\right)^2.
\end{equation}
This expression may be simplified in the limit of large $A_0$ if the following 
integrals over all $R^2$ converge
\begin{eqnarray}
v_0 &=& \int d{\bf x}\  V(|{\bf x}|) = 2\pi \int_0^\infty dr\  r V(r)\\
v_1 &=& \int d{\bf x}\ {V'(|{\bf x}|)\over |{\bf x}|} = 2\pi \int_0^\infty dr
V'(r) = 2\pi (V(\infty) -V(0)).\label{eqv0}
\end{eqnarray}
The above integrals converge when $V(0)$ is finite, which can be ensured 
via a suitable regularization scheme, and when $V(r)$ tends to zero quicker
than $1/r^2$ for large $r$.
In this case we find 
\begin{equation}
H = {1\over 2}\int_{A_0 \times A_0} \ d{\bf x} d{\bf x' }
h({\bf x}) G({\bf x-x'}) h({\bf x'}),
\end{equation}
where the operator $G$ is given by
\begin{equation}
G({\bf x}) = -{v_0\over 2}\nabla^2 \delta({\bf x}) +v_1 \delta({\bf x})
-  {V'(|{\bf x}|)\over |{\bf x}|}.
\end{equation}
The two dimensional Fourier transform of $G$ is given by
\begin{equation}
{\tilde G}(q) = v_1 +{v_0 q^2 \over 2} -2\pi \int_0^{\infty}  dr
V'(r)J_0(qr)
\end{equation}
where
\begin{equation}
J_0(qr) = {1\over 2\pi}\int_0^{2\pi} d\theta \exp\left(iqr\cos(\theta)\right)
\end{equation}
is a Bessel function of the first kind \cite{abram}. 
This results may be written as
\begin{equation}
 {\tilde G}(q) = 2\pi \int_0^\infty dr\left( 1 - {q^2r^2\over 4} 
-J_0(qr)\right)
 V'(r),\label{eqnetz1}
\end{equation}
recovering the result of Netz \cite{netz} by using the  identities
\begin{equation}
2\pi \int_0^\infty dr V'(r)  = v_1
\end{equation}
and 
\begin{equation}
\pi \int_0^\infty dr r^2 V'(r)  = -v_0,
\end{equation}
where the second assumption is valid providing that 
$\lim_{r\to\infty} V(r)r^2 = 0$. An integration by parts can be used to rewrite
Eq. (\ref{eqnetz1}) as \cite{netz}
\begin{equation}
 {\tilde G}(q) = 2\pi \int_0^\infty dr\left( q^2 r -2 qJ_1(qr)\right)
 V(r).\label{eqnetz2}
\end{equation}
The small $q$ expansion of  Eq. (\ref{eqnetz2}) gives
\begin{equation}
{\tilde G}(q) = {\pi q^4 \over 8}\int_0^\infty dr V(r)r^3 +O(q^6).
\end{equation} 
Going back to real space this low momentum term gives a contribution to
the Helfrich energy of
\begin{equation}
\Delta_{Hel} = {\delta \kappa_b\over 2}\int_{A_0\times A_0} d{\bf x} \left(\nabla^2 h({\bf x})\right)^2
\end{equation}
where 
\begin{equation}
\delta\kappa_b ={\pi\over 8}\int_0^\infty dr V(r)r^3,
\end{equation}
is the renormalization of the membrane bending rigidity. 
\section{General calculation for  cylinders and spheres}
In a cylindrical geometry the energy for an interaction $V$ between
membrane components with the corresponding energy of a flat membrane of
the same area subtracted off is the bending energy. For sufficiently
short range interactions it is given by
\begin{equation}
H_b = {2 \pi R L\over 2}\left[
\int_{-\infty}^{\infty} dz \ \int_{-\pi}^{\pi} R d\theta
V\left(\sqrt{4R^2 \sin^2({\theta\over 2}) + z^2}\right)
- 2\pi \int_0^{\infty} dr\ r V(r)\right]
\end{equation}
The fact that the potential is short ranged is used in setting the limits of the $z$ integration in the first integral at $\pm \infty$ and the limits of 
the $r$ integration in the second integral (for the flat membrane) at 
$0$ and $\infty$. In the first integral the only contribution that will
be present for a short range potential will be from the region $\theta = 0$,
more specifically the region where $\theta = \alpha/R$ and where $\alpha \sim O(1)$. Making this change of variables and expanding the $\sin$ we obtain the 
first integral in the above to be
\begin{equation}
I = \int_{-\infty}^{\infty} dz \ \int_{-\pi R}^{\pi R }d\alpha  
V\left(\sqrt{\alpha^2 - {\alpha^4\over R^2} +O(1/R^4) + z^2}\right)
\end{equation}
To leading order in $1/R$ this is
\begin{equation}
I =  \int_{-\infty}^{\infty} dz \ \int_{-\pi R}^{\pi R } d\alpha
\left[ V(\sqrt{\alpha^2 + z^2}) - {\alpha^4\over 24 R^2 \sqrt{\alpha^2 + z^2}}
V'(\sqrt{\alpha^2 + z^2})\right].
\end{equation}
For large $R$ this can be written as
\begin{equation}
I =  \int_{0}^{\infty} dr \int_0^{2\pi}d\theta 
\left[r V(r) - {r^4 \cos^4(\theta)\over 24 R^2} V'(r)\right],
\end{equation}
where we have written $\alpha = r\cos(\theta)$ and $z = r\sin(\theta)$.
We thus find that to leading order 
\begin{equation}
H_b = {\pi \delta\kappa_b L \over R}
\end{equation}
with 
\begin{eqnarray}
\delta\kappa_b  &=& -{1\over 24}\int_0^{2\pi} d\theta  \cos^4(\theta)\int_0^{\infty}
dr \ r^4 V'(r) \nonumber \\
&=& {\pi\over 8} \int_0^{\infty}
dr \ r^3 V(r).\label{kcyl}
\end{eqnarray}
Therefore we see that the  calculation in an explicitly cylindrical geometry 
agrees with that
obtained by the formalism of Netz. This makes it unlikely that 
the disagreement between \cite{netz} and \cite{pinc} is due to a
difference of topology, at least for suitably short range potentials.
However the long-range nature of the potential may play a crucial role
in the case of a salty membrane with no external electrolyte. In the 
calculations presented above one sees clearly that the first order 
contribution to the bending energy comes from local curvature. This
means that the energy associated with a given point comes from the 
local curvature about that point. For a sphere, for example, one is
clearly neglecting the contributions from the sphere which come 
from the interactions with points on the opposite side. If the potential
$V$ has no intrinsic scale and has a power law form 
$V(r) \sim 1/r^\alpha$  at large $r$, then the contribution from a point
due to points opposite to it is of the order $R^{2-\alpha}$ where the term 
$R^2$ is the area term. This means that this contribution is negligible
for $\alpha >2$, but plays an role for $\alpha \leq 2$. Indeed one
sees that the first order term in the energy of the system per unit area, 
which corresponds to the flat membrane is given by
$2\pi \int_0^\infty r dr V(r)$ and this is clearly divergent when 
$\alpha \leq 2$. Furthermore it is clear from Eq. (\ref{eqv0}) that even
the analysis about the flat membrane breaks down in the case $\alpha <2$.

Let us now consider the same calculation but for a sphere. For 
suitably short range potentials the bending energy of the sphere is given by

\begin{eqnarray}
H_b &=& 2\pi R^2 \left[R^2\int_0^\pi \sin(\theta) d\theta
\int_0^{2\pi} d\phi\  V\left( R \sqrt{2- 2\cos(\theta)}\right)
-2\pi \int_0^{\infty} dr \ r V(r)\right]\nonumber \\
&=& 4\pi^2 R^2 \left[\int_0^{2R} dr \ rV(r) -  \int_0^{\infty} dr \ rV(r)\right]
\nonumber \\
&=&
-4\pi^2 R^2 \int_{2R}^{\infty} dr \ r V(r),\label{eqsph}
\end{eqnarray}
where to obtain the above we have made the change of variables
$ 2-2\cos(\theta) = r^2/R^2$. We see from Eq. (\ref{eqsph}) that if the 
potential $V$ decays exponentially, or faster than exponentially,  
as a function of $R$, then the bending energy has an 
exponential, or faster than exponential, decay. For a 
sphere the bending energy takes the form
\begin{equation}
H_b = 8\pi \kappa_b + 4\pi \kappa_g,
\end{equation}
where $\kappa_b$ is the bending rigidity and $\kappa_g$ is the Gaussian
rigidity \cite{boal}. For short range potentials the above calculation
implies that the presence of a long, but finite range, interaction does not
renormalize the bending energy of a sphere to first order, rather we 
conclude that $\delta \kappa_g = -2 \delta \kappa_b$ -- 
so the renormalization of the
Gaussian rigidity is of the opposite sign to that of the bending rigidity.
We note that the signs of these results are in agreement with the 
results on charged membranes
\cite{winh,kum}. 
It is interesting to note the
calculations based on perturbations about a flat plane do not indicate
any renormalization of the Gaussian  rigidity by long-range 
interactions. This must be due to the inherently topological nature
of the Gaussian bending energy \cite{boal}, which is insensitive to
the geometry of the system and only depends on the membrane topology. It
is thus normal that this energy is not picked up by a local perturbative
analysis. Notice that if $V$ is negative, and decaying at $R\to \infty$
sufficiently quickly, then the bending energy is actually positive -
at variance with the intuition that attractive interactions favor bending !

\section{Membranes of finite thickness}

In reality the membrane will always have a finite thickness.
In the case of a cylinder we will examine the effect of long-range interactions
on the bending rigidity of a cylindrical shell $C$ of uniform thickness 
$\delta$. We will take the outer radius of the cylinder to be at $R+\delta/2$ 
and the internal radius at $R+\delta/2$. If the cylinder is of length $L$
then the volume of the shell is $2\pi RL\delta$. The energy of the cylinder
is now given by
\begin{equation}
H = {1\over 2}\int_{C\times C} d{\bf r}_1 d{\bf r}_2 V\left(|{\bf r}_1 -{\bf r}_2|\right),
\end{equation} 
Note that the potential $V$ as defined above is now due to the interaction 
of volumes and not areas, it thus has physical dimensions $[E]/[L^6]$.
The energy above can be written as
\begin{equation}
H = \pi L \int r_1 r_2 dr_1 dr_2 d\theta dz
V\left(\sqrt{(r_1-r_2)^2 + 2 r_1 r_2 (1-\cos(\theta)) + z^2}\right),
\end{equation}
where in the above the integration ranges are $(R-\delta/2, R+\delta/2)$ for
$r_1$ and $r_2$, $(-\pi,\pi)$ for $\theta$ and $(-\infty,\infty)$ for $z$.
One now makes the change of variables $\theta = \alpha/\sqrt{r_1 r_2}$,
and recalling that $r_1$ and $r_2$ are of order $R$, expand the argument
of $V$ to fourth order in $\alpha$ to obtain
\begin{equation}
H = \pi L \int \sqrt{r_1 r_2}  dr_1 dr_2  dz d\alpha \left[ 
V\left(\sqrt{(r_1-r_2)^2 + z^2 +\alpha^2 }\right)
- {\alpha^4 \over 24 r_1 r_2} {V'\left( \sqrt{(r_1-r_2)^2 + z^2 +\alpha^2 }\right)\over \sqrt{(r_1-r_2)^2 + z^2 +\alpha^2 }}\right].
\end{equation}
Again taking the limits of the $\alpha$ integrations to $\pm \infty$, carrying out the $\theta$ integration, then replacing the coordinates $(\alpha,z)$ by the radial coordinates $(r,\theta')$ and writing $r_i = R+x_i$ for $i=1$ and $2$
we find that to order $1/R$
\begin{eqnarray}
H = &\pi& L R \int dx_1 dx_2 r dr d\theta' \left[
(1 + {1\over 2R}(x_1 + x_2) + {1\over 2 R^2}x_1 x_2 -{1\over 8 R^2}
(x_1 + x_2)^2)V\left(\sqrt{(x_1-x_2)^2 + r^2}\right)\right. \nonumber \\
&-& \left.{1\over 24 R^2} r^3 \cos^4(\theta') {V'\left(\sqrt{(x_1-x_2)^2 + 
r^2}\right)\over \sqrt{(x_1-x_2)^2 + r^2}}
\right]\label{eqstep}
\end{eqnarray}
Now carrying out the $\theta'$ integration and an integration by 
parts on the last term above we obtain.

\begin{equation}
H = H_{bulk} + H_b
\end{equation}
where $H_{bulk}$ is the bulk energy dependent only on the volume and given by
\begin{equation}
H_{bulk} = 2\pi^2 LR \int dx_1 dx_2 r dr \ 
V\left(\sqrt{(x_1-x_2)^2 + r^2}\right),
\end{equation}
and $H_b$ is the bending energy given by
\begin{equation}
H_b ={\pi^2 L\over 8 R}\int dx_1 dx_2 r dr \left(r^2 - 2(x_1-x_2)^2\right)
V\left(\sqrt{(x_1-x_2)^2 + r^2}\right).
\end{equation}
Note that in the above, the $x_1$ and $x_2$ integrations are
over  $(-\delta/2,\delta/2)$. The above result can be simplified 
slightly by changing variables and writing $w = x_1-x_2$ and $u = x_2 + x_1$,
and noting that for $w$ positive the integration range for the variable 
$u$ is then over $(-\delta+w,\delta-w)$. Therefore 
for a generic function $f(x,y)$ even in both its arguments we have
\begin{equation}
\int dx_1dx_2 f(x_1-x_2,x_1+x_2) = 
\int_0^\delta dw\int_{-\delta+w}^{\delta-w}du\ f(w,u).
\end{equation}
Using this change of variables we then obtain
\begin{equation}
H_b ={\pi^2 L\delta \over 4 R}\int_0^\delta dw\ (\delta-w)\int_0^\infty r dr 
\left(r^2 - 2w^2\right)
V\left(\sqrt{w^2 + r^2}\right)\label{hthick}.
\end{equation}
If we take the limit of small $\delta$ we obtain
\begin{equation}
H_b = {\delta^2 \pi^2 L\over 8 R}\int  r^3 dr\ V(r),\label{eqhh}
\end{equation}
{\em provided} the above integral is finite and providing that
the integration by parts carried out on the last term of 
Eq. (\ref{eqstep}) is valid. 
We note that this result is 
in agreement  with Eq. (\ref{kcyl}) as the effective interaction between 
unit areas  for thin shells, {\em i.e.} for small $\delta$, is $\delta^2 V$.

The result Eq. (\ref{hthick}) yields a bending rigidity (here we explicitly
include all the integration limits)
\begin{equation}
\kappa_b = {\pi \delta \over 8 }\int_0^\delta dw\ (\delta-w) \int_0^\infty 
r dr \left(r^2 - 2w^2\right)
V\left(\sqrt{w^2 + r^2}\right)\label{khthick}.
\end{equation}

The same calculation can be carried out for a sphere and one finds that the 
total energy is given by
\begin{equation}
H = 4\pi ^2 \int r_1^2 dr_1\ r_2^2 dr_2 \ \sin(\theta) d\theta\ V\left(\sqrt{(r_1-r_2)^2 +
2r_1 r_2 (1-\cos(\theta))}\right)
\end{equation}
where the $r_1$ and $r_2$ integrations are over $(R-\delta/2,R+\delta/2)$.
One now makes the change of variables
\begin{equation}
r^2 = 2r_1r_2\left(1- \cos(\theta)\right)
\end{equation}
which gives
\begin{equation}
H= \pi^2 \int r dr dr_1 dr_2 \left[ (r_1+r_2)^2 -(r_1-r_2)^2\right] V\left(
\sqrt{(r_1-r_2)^2 + r^2}\right),
\end{equation}
where for large $R$ the limits of the $r$ integration can be taken to be
$(0,\infty)$. Now writing $r_1= R+ x_1$ and $r_2 = R+x_2$ then switching to the
variables $u$ and $w$ as above we find
\begin{equation}
H = \pi^2\int_0^\delta dw\ \int_{-\delta+w}^{\delta-w}du\ 
\left[ (2R+u)^2 -w^2\right]   \int_0^\infty rdr\ V\left(
\sqrt{ r^2+w^2}\right).
\end{equation}
Now carrying out the $u$ integral yields
\begin{eqnarray}
H &=& 8\pi^2 R^2 \int_0^\delta dw (\delta -w)\int_0^\infty rdr\ 
V\left(
\sqrt{ r^2+w^2}\right) \nonumber \\
&+& \pi^2  \int_0^\delta dw (\delta -w)[ {2\over 3}\delta^2 -{4\over 3}\delta w -{4\over 3} w^2] \int_0^\infty rdr\ V\left(\sqrt{ r^2+w^2}\right).
\end{eqnarray}
Now to obtain the bending energy we must subtract the bulk energy from
this result. The area of the flat membrane necessary to form this sphere
is of radius given by
\begin{eqnarray}
A_S(R) &=& {4\pi \over 3\delta}\left((R+{\delta\over 2})^3 - 
(R-{\delta\over 2})^3\right)\nonumber \\
&=& 4\pi R^2 + {\pi\delta^2\over 3}
\end{eqnarray}
The energy of a flat membrane of this area is then given by
\begin{eqnarray}
H_{bulk} &=& {1\over 2}\times  A_S(R)\times 
2\pi\int_0^\delta dw 2(\delta-w) \int_0^\infty rdr \ V\left(\sqrt{ r^2+w^2}\right)
\nonumber \\
&=& \left[ 8\pi^2 R^2 + {2\pi^2\delta^2\over 3}\right]\int_0^\delta dw (\delta-w) \int_0^\infty rdr \ V\left(\sqrt{ r^2+w^2}\right) 
\end{eqnarray}
In calculating $H_b = H-H_{bulk}$ we see that the term proportional to the
surface area cancels and we are left with
\begin{equation}
H_b = -{4\pi^2 \over 3} \int_0^\delta dw \ w(\delta^2 -w^2)\int_0^\infty rdr \ V\left(\sqrt{ r^2+w^2}\right),\label{sthick}
\end{equation}
which thus implies
\begin{equation}
2\kappa_b +\kappa_g = -{\pi \over 3} \int_0^\delta dw \ w(\delta^2 -w^2)
\int_0^\infty rdr \ V\left(\sqrt{ r^2+w^2}\right)\label{spsum}
\end{equation}        

The result Eq. (\ref{sthick}) is rather significant as it shows that a 
spherical vesicle has a non-zero bending energy at large $R$  
when  one takes into account the finite thickness of the membrane. 
When $\delta$ is small and the corresponding integrals turn out to be finite
we may ignore the $w$ dependence in the argument of $V$ in  Eq. (\ref{spsum})
to obtain a formula analogous to Eq. (\ref{eqhh}):
\begin{equation}
2\kappa_b +\kappa_g = -{\pi \delta^4\over 12}\int_0^\infty rdr \ V(r).
\end{equation}
\section{Effect of  diffuse van der Waals interactions}

Here we consider the problem where the membrane of finite thickness
has a different dielectric constant to that of the external
media/solvent . We take the dielectric constant of the membrane to be 
$\epsilon'$ while the external medium has dielectric constant $\epsilon$. 
This difference in 
dielectric gives rise to a thermal Casimir effect which is  a static 
van der Waals interaction. In general these interactions are not pairwise,
but in the diffuse limit, where $\epsilon -\epsilon'$ is small, the pairwise
part of the interaction is  the dominant one. 

For a system  volume $C$ having dielectric constant $\epsilon'$ in an external
medium of dielectric constant $\epsilon$, the partition function for the 
thermal fluctuations of the zero frequency Matsubara modes of the 
electrostatic field is given by \cite{lif,mani,deho2,netz2}
\begin{equation}
Z= \int d[\phi] \exp\left(S_0 + \Delta S\right)\label{eqZ}
\end{equation}
where 
\begin{equation} 
S_0 = -{\epsilon\over 2}\int d{\bf x} (\nabla \phi)^2
\end{equation}
and 
\begin{equation}
\Delta S = -{\epsilon'-\epsilon\over 2}\int_C  d{\bf x} (\nabla \phi)^2.
\end{equation}
Note that in $S$ the integral is over all space but in $\Delta S$ the integral
is only over the volume $C$ containing the media of differing dielectric 
constant to the exterior. As the action in Eq. (\ref{eqZ}) is quadratic
the partition function can be written as a functional determinant, however
for general geometries this calculation is rather complicated. 
In the case where $\Delta S$ is
small one may carry out a cumulant expansion to second order which gives
for the free energy $F$ of the system
\begin{equation}
F - F_0 = {-k_B T} \left( \langle \Delta S\rangle_0 +
{1\over 2}  \langle \Delta S^2\rangle_{0,c}\right).\label{eqVE}
\end{equation}
Here $F_0$ is simply the vacuum free energy in the absence of the membrane 
$C$. The subscript $0$ indicates that the expectation is taken with 
respect to the vacuum measure $S_0$ and thus 
\begin{equation}
\langle \phi({\bf x})\phi({\bf y})\rangle_0 = G_0({\bf x} -{\bf y}),
\end{equation}
where $G_0$ is the Green's function of the free theory with action $S_0$
and the subscript $c$ denotes the connected part.
Now we note that the first term of Eq. (\ref{eqVE}) is simply a
term proportional to the volume of the system and thus does not contribute
to the bending free energy. The dilute approximation to the bending free
energy is thus
\begin{equation}
H =  -{k_B T \over 2}  \langle \Delta S^2\rangle_{0,c},
\end{equation}
In general if one writes
\begin{equation}
\Delta S = \int_C d{\bf x} \ R[\phi({\bf x})],
\end{equation}
then we obtain
\begin{equation}
H =  -{k_B T \over 2}\int_C  d{\bf x}
d{\bf y} \langle R[\phi({\bf x})]R[\phi({\bf y})]\rangle_{0,c},
\end{equation}
which means that the effective pairwise potential $V$ is given by
\begin{equation}
V({\bf x}) = -k_B T \langle R[\phi({\bf x}]R[\phi({\bf 0})]\rangle_{0,c}
\end{equation}
In the case of a difference in dielectric constants we thus obtain the potential\begin{equation}
V({\bf x}) = -k_B T 
{(\epsilon -\epsilon')^2\over 4}\beta^2\langle (\nabla \phi({\bf x}))^2 (\nabla \phi({\bf 0}))^2\rangle_{0,c}
\end{equation}
Now using the fact that $G_0({\bf x}) = G_0(r)$ where $r = |{\bf x}|$ 
we obtain
\begin{equation}
\langle (\nabla \phi({\bf x}))^2 (\nabla \phi({\bf 0}))^2\rangle_{0,c}
= {4\over r^2} \left({dG_0\over dr}\right)^2 + 
2\left({d^2G_0\over d^2r}\right)^2
\end{equation}
Here the Green's function $G_0$ is given by
\begin{equation}
G_0 ({\bf x})= -{1\over 4\pi \epsilon \beta |{\bf x}|}
\end{equation}
and thus we obtain
\begin{equation}
V({\bf x}) = -{A\over |{\bf x}|^6}
\end{equation}
where
\begin{eqnarray}
A&=& 3k_B T 
{(\epsilon -\epsilon')^2\over 16 \pi^2 \epsilon^2}\nonumber \\
&\approx & 
3k_B T {\Delta^2\over 4 \pi^2},
\end{eqnarray}
where $\Delta = (\epsilon -\epsilon')/2\epsilon\approx 
(\epsilon -\epsilon')/(\epsilon +\epsilon')$ to leading order in 
$\epsilon -\epsilon'$. This, as is to be expected, recovers the 
standard form of the unretarded van der Waals interaction.  

We now examine the bending rigidity of a membrane 
of finite thickness $\delta$ with this interaction. We 
regularize the interaction by writing
\begin{equation}
V({\bf x}) = -{A\over ({\bf x}^2 + a^2)^3},
\end{equation}
where $a$ is a short-scale cutoff. Substituting this into Eq. (\ref{khthick})
the $r$ integration is easily performed to yield, after some algebra,
\begin{equation}
\kappa_b = -{3 k_B T\Delta^2 \over 64 \pi}\int_0^\delta dw \ (\delta-w)\left[ {2a^2\over( a^2 +
w^2)^2} -{1\over a^2 +  w^2}\right]
\end{equation}
which then gives
\begin{equation}
 \kappa_b = -{3 k_B T\Delta^2 \over  128 \pi}\ln\left({\delta^2 + a^2\over
a^2}\right).
\label{kbvdw}
\end{equation}
We see that, as with the eigen-mode expansion method of \cite{deho}, 
for small cut-off $a$ the bending rigidity has the form
\begin{equation}
\kappa_b = -{3 k_B T\Delta^2\over  64 \pi}\ln({\delta\over a}),
\end{equation}
and that the above result is identical to  Eq. (\ref{vdw1}) up to a rescaling
of the microscopic cut-off $a$.

The calculation for the sphere is also straight forward to carry out and one
obtains
\begin{equation}
\kappa_g + 2\kappa_b = {k_B T\Delta^2\over 32 \pi}
\left[{\delta^2\over a^2} -\ln\left({\delta^2 + a^2\over
a^2}\right)\right]\label{eqsst}
\end{equation}
which along with Eq. (\ref{kbvdw}) gives
\begin{equation}
\kappa_g  = {k_B T\Delta^2\over 64 \pi}
\left[2{\delta^2\over a^2} +\ln\left({\delta^2 + a^2\over
a^2}\right)\right] \label{kgvdw}.
\end{equation}
Note that from Eq. (\ref{eqsst}) total bending energy for the sphere
is positive for all values of $\delta/a$.  
We see from Eq. (\ref{kbvdw}) that the renormalization of $\kappa_b$ 
due to static van der Waals interactions is rather weak. Although the result
depends on the short scale cut-off it does so only logarithmically. 
Physically realistic values of $\delta$ and $a$, corresponding to the membrane 
thickness and a typical dipole size or dipole separation gives at most
an $O(k_B T)$ renormalization of $\kappa_b$  \cite{deho}. 
This is much smaller than the experimental  values obtained for 
$\kappa_b$ which tend to be between $3-30\  k_B T$ \cite{boal}. 
However Eq. (\ref{kgvdw}) 
predicts a $\kappa_g$ which depends strongly on $\delta$ and it is 
conceivable that van der Waals interactions make a significant 
contribution to $\kappa_g$. Unfortunately few experimental measurements 
or estimates exist for $\kappa_g$. 
\section{Conclusions}
In this paper we have revisited the problem of the renormalization of
the bending and Gaussian rigidities of membranes by long-range interactions.
These renormalizations may be calculated via a geometric approach 
applied to cylindrical and  spherical geometries. The result obtained for
$\kappa_b$ is found to agree with that found for a perturbative analysis
about a flat membrane in a general approach proposed by Netz \cite{netz}. 
We also rederived the Netz result for a flat membrane by considering an
ensemble where the membrane is thought of as being drawn from a reservoir
of flat non-fluctuating membrane showing the equivalence of the two
approaches from an ensemble point of view.
Using the geometric approach we obtained the
general result for an infinitesimally thin membrane that $\kappa_g = -
2 \kappa_b$, {\em i.e.} the somewhat surprising result that the Gaussian
rigidity is renormalized by a long-range potential with a sign opposite
to the bending rigidity. This effect has also been seen
in more specific mean-field studies of charged membranes. 
We then derived analogous formulas for the 
bending and Gaussian rigidity when the membrane has a finite (but small
relative to the radii of curvature) thickness. Finally we calculated the 
bending and Gaussian rigidities induced by the thermal Casimir force, or 
static van der Waals interactions, for a dilute system. In agreement with 
our previous studies we find a negative contribution to the bending
rigidity but with exactly the same functional dependence on the thickness
$\delta$ of the membrane and the microscopic cut-off $a$ when $a$ is small
compared to $\delta$. This agreement is reassuring as although  microscopic
details  are dominating the physics they are doing so in a rather universal
way which is insensitive to the regularization scheme being employed.  
The Gaussian rigidity is found to undergo a positive renormalization 
due to van  der Waals interaction. This renormalization  strongly divergent  
as $a\to 0$ behaving as $\delta^2/a^2$. As pointed out in 
\cite{deho}, van der Waals interactions only weakly favor  the 
formation of tube like structures, such as t-tubules and 
it is unlikely that they can stabilize cylindrical geometries 
thermodynamically. However if  these 
structures are formed via another physical or  biological mechanism, 
then attractive van der  Waals forces may contribute to their stability 
in that they will impede the formation of spherical budding instabilities 
and thus enhance the metastability of these structures. Whether this
enhancement of metastability is significant depends  the value of 
$\delta / a $ and contributions of a similar functional form which will
come from the non-zero frequency frequency Matsubara modes which are
responsible for the  fluctuating part of the van der Waals interaction.

A final comment on the conflicting results 
Eq. (\ref{spincus}) and Eq. (\ref{snetz}) for the renormalization of the 
bending rigidity of a salty membrane is in order. In the case of a
system with salt outside, the bare interaction induced between the 
membrane components, in the diffuse limit where the salt concentration 
within the membrane is small, behaves as $\exp(-2mr)/r^2$ where $m$ is the 
external Debye mass. In this limit the two results should give the same
result as the interaction is sufficiently short range. In the Netz approach
an effective pairwise interaction is computed, however in the eigen-mode
expansion of Pincus and Lau it is clear from their calculation that they are
summing all terms in the eigen-mode expansion and not just those which
correspond to the dilute-pairwise limit. Concretely the logarithms in the 
expressions given in \cite{pinc} are not expanded to second order. 
This means that their result is inherently taking into account 
multiple scattering events and thus non-pairwise interactions. 
All the same one should bare in mind that the physical situation 
where one has salt in the membrane but non outside is rather unlikely.

\end{document}